\documentclass{iau}

\usepackage{amsmath}
\usepackage{graphicx}
\usepackage{multirow}

\begin{document}

\lefttitle{Andreas A.C. Sander}
\righttitle{Open Questions in Massive Star Research across Cosmic Scales}

\jnlPage{1}{7}
\jnlDoiYr{2025}
\volno{402}
\doival{}

\aopheadtitle{Proceedings IAU Symposium}
\editors{A. Wofford,  N. St-Louis, M. Garcia \&  S. Simón-Díaz, eds.}

\title{Open Questions in Massive Star Research across Cosmic Scales}

\author{Andreas A.C. Sander}
\affiliation{{Zentrum f{\"u}r Astronomie der Universit{\"a}t Heidelberg, Astronomisches Rechen-Institut, M{\"o}nchhofstr. 12-14, 69120 Heidelberg, Germany}}

\begin{abstract}
Massive stars are the engines of the Cosmos, shaping their environments and driving galaxy evolution across cosmic time. Yet, this general textbook picture faces many challenges when trying to turn abstract insights into quantitative predictions. Recent discoveries, such the surprisingly high metallicity and early nitrogen enrichment in high-redshift galaxies discovered by JWST, are challenging current descriptions of massive star evolution and add new pieces to a puzzle that is yet everything but complete.

The oncoming era of large surveys and advances in computational modeling create the potential to reach breakthroughs in our understanding. Yet, to resolve current problems and conflicting conclusions, we will also need to reconsider what we think we know. Are the objects we observe what we think they are? Are the models we use describing what is actually going on? And what can we learn from previous misconceptions? This short review highlights major open questions from individual stars and stellar systems back to the first galaxies while also
discussing two examples -- the weak-wind problem as well as the different flavours and impact of Wolf-Rayet stars -- where recent discoveries might point in a new direction.
\end{abstract}


\maketitle

\section{Open Questions across cosmic scales}

Massive Stars are shaping our Universe ever since the first generation of stars. Spending most of their lifetime as hot stars, massive stars ionize their surrounding environment and have been a key factor in making the Universe transparent. In their interiors, nuclear fusion generates the elements that are the basics of nucleosynthesis and our periodic table. Via internal mixing and eventual mass loss from the surface as well as powerful supernovae explosions accompanying many of the eventual core collapse events, massive stars then not only distribute these elements into space, but also inject kinetic energy and momentum via their winds, outbursts, or companion interactions. Thus, with each generation of massive stars, the average amount of elements heavier than helium, in astrophysics simply termed ``metals'', increases. This ``metallicity'' $Z$ can thus act as an important proxy of cosmic time. 

Yet, our textbook picture of massive stars requires updates on many pages as we struggle to get a coherent understanding, quantitatively aligned with observations and theoretical predictions. Especially when moving accross traditional field boundaries, we find both conflicting descriptions as well as unphysical but strangely ``sufficient'' treaments. A prominent example for the former are the sizes of clumps in stellar winds, which should be tiny to explain the overall constant appearance of a stellar spectrum, but huge when accreted onto a compact object nearby to explain X-ray lightcurve variations. On larger scales, single star evolution modelling seems to be remarkably sufficient to reproduce the appearance of star-forming galaxies despite the high multiplicity fraction in the massive star regime and the failure to reproduce many observed individual stars and systems. Trying to resolve these conflicting results quickly leads back to questioning basics we thought to be settled. Hence, what might sound like a problem in the details at first actually hints at big open questions and the search for larger, conflict-free framework for massive stars which is much needed, but so far missing.

\begin{figure}[t]
\includegraphics[width=\textwidth]{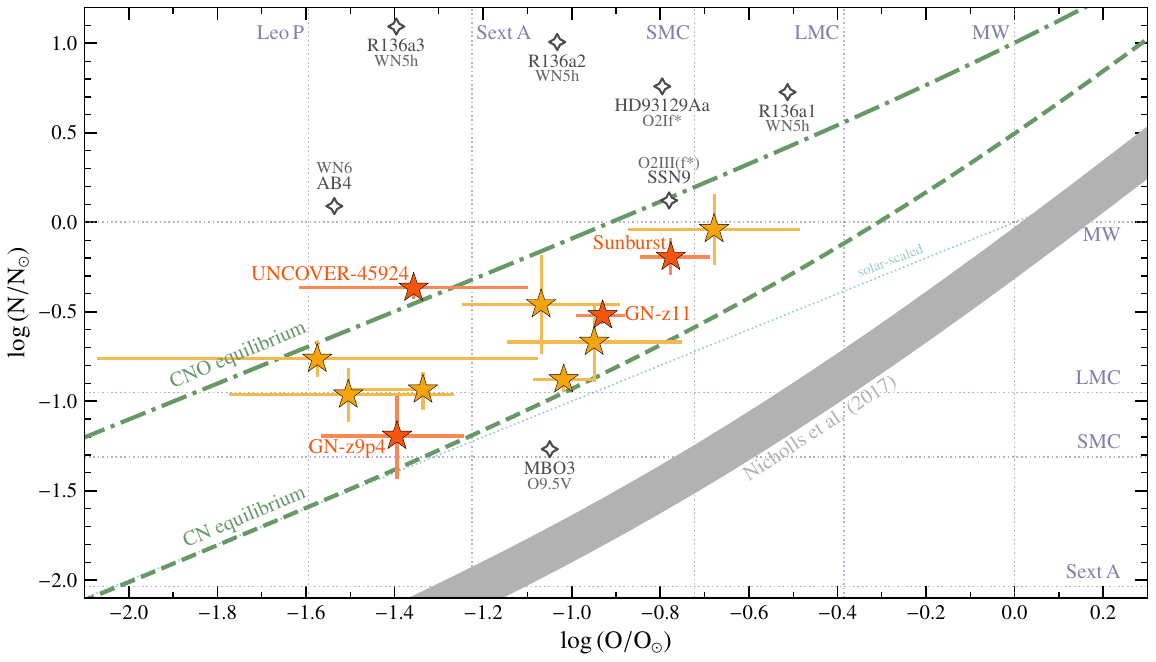}
\caption{Galaxies with high nitrogen enrichment \citep[yellow/orange stars, taken from][]{Pascale+2023,Castellano+2024,Marques-Chaves+2024,Schaerer+2024,Senchyna+2024,Topping+2024,Ji+2025,Napolitano+2025} compared to typical abundances in star-forming regions \citep[gray area,][]{Nicholls+2017} in the C-O-mass-fraction plane, normalized to solar abundances from \citet{Magg+2022}. Green curves show CN (dashed) and CNO (dashed-dotted) equilibrium abundances \citep[assuming scaling by][]{Nicholls+2017}. For comparison, abundances of few individual stars (R136: \citealt{Brands+2022}; HD93129Aa: \citealt{Gruner+2019}; SSN9: \citealt{Bouret+2021}; AB4: \citealt{Pauli+2023}; MBO3: \citealt{Ramachandran+2021}) from different environments are shown as black, 4-pointed stars. Baseline abundances (purple dotted lines) are from \citet{Kniazev+2005}, \citet{Skillman+2013}, and \citet{Vink+2023}.}
\label{fig:enrichment}
\end{figure}

\subsection{Puzzling discoveries at high redshift}

With the unique infrared capabilities of the James Webb Space Telescope (JWST), a new door has opened to observe galaxies soon after the dark ages. Yet, the spectra obtained from star-forming galaxies have led to multiple puzzling discoveries that do not fit in our (previous) view about massive stars, most prominently reionization-era galaxies highly enriched in nitrogen (N). Fig.\,\ref{fig:enrichment} shows a sample of these objects in the N versus oxygen (O) plane. The galaxies are significantly more N-rich than most known star-forming galaxies and H\,II regions, but all targets essentially are located between the CN and CNO equilibrium which could be reached even if the galaxies are normal in C/O. Individual massive stars can reach even more extreme surface abundances as illustrated by a few examples in Fig.\,\ref{fig:enrichment}. Lots of open questions around these findings are currently discussed \citep[e.g.,][]{Cameron+2023,Marques-Chaves+2024,Berg+2025,Arellano-Cordova+2025,Morel+2025}:
Could more efficient and bursty star-formation be responsible? As the high-redshift galaxies are extremely dense, do we maybe see the progenitors of Globular Clusters, possibly sparked by extremely massive stars or even supermassive stars? Or are massive and very massive stars sufficient and the problem in our understanding is rather located in the diagnostics and well as the evolution and feedback models? In particular the latter is not getting easier by the fact that metallicity measurements at high redshift are essentially all O-based, while population synthesis models include assumptions about stellar evolution that are significantly depending on the iron abundance. In parallel, signatures of highly ionized environments are also found in many low-metallicity and high-redshift galaxies. While their origin is still quite unclear, some possible sources are listed in Table\,\ref{tab:ion} and will be discussed further below.

	  \begin{table}
			 \centering
			 \caption{\label{tab:binmis} Selected examples of objects that got significant parameter revisions after performing a detailed quantitative spectroscopic analysis and revealing multiplicity }
				\begin{tabular}{llcll} 
			    AzV 476     &  O2--3 V                     & $\rightarrow$       &  O4 IV-III ((f))p + O9.5: Vn  & \citep{Pauli+2022} \\ %
                      & \small $> 45\,$kK            & \small$\rightarrow$ &  \small$42\,$kK + $32\,$kK &             \\ 
					            & \small $\gtrsim 60\,M_\odot$ & \small$\rightarrow$ &  \small$20\,M_\odot$ + $18\,M_\odot$ &   \\[0.5em]     
	    		2dFS 163    &  O8 Ib (f)                   & $\rightarrow$       &  O6.5 Ib (f) + B0-1 Ve        &  \citep{Ramachandran+2024} \\ %
		                  & \small $\sim$$33\,$kK        & \small$\rightarrow$ &  \small$37\,$kK + $26\,$kK  \\
					            & \small $\sim$$37\,M_\odot$   & \small$\rightarrow$ &  \small$4\,M_\odot$ + $11\,M_\odot$ &   \\     
			 \end{tabular}
%			    SSN 7       &   O4 III(n)(f)              & $\rightarrow$       &   ON3 If*+O5.5 V((f))    & \citep{RickardPauli2023} \\
%                     & \small $\sim$$42\,$kK       & \small$\rightarrow$ &  \small$44\,$kK + $39\,$kK &             \\ 
%                     & \small $\sim$$49\,M_\odot$  & \small$\rightarrow$ &  \small$32\,M_\odot$ + $55\,M_\odot$ &   \\   (Algol-like system)
	  \end{table}

\subsection{The need to connect the scales and uncover multiplicity}

  While yielding fascinating new discoveries, high-redshift observations alone will be insufficient to answer any of these questions robustly as the resolution and wavelength regimes are quite limited. A whole ladder of local analogs and testbed environments is required to study and connect the involved physical processes, ranging from nearby (dwarf) galaxies with lower $Z$ over unresolved and resolved clusters to individual stellar systems such as the iron-poor O stars in the Magellanic Bridge \citep{Schoesser+2025}. Resolved clusters are a particular important step as individual stars can be studied in detail while the cluster environment yields additional constraints and provides a testbed for population synthesis. 
So far, few clusters have been studied with detailed, multi-epoch quantitative spectroscopy with the LMC Tarantula region marking the most prominent example. Further clusters now coming more into the focus are Westerlund 1 (Wd1) and Per OB1 in the Milky Way and NGC\,346 in the SMC. 
	
Within clusters or in the field, resolving and understanding the multiplicity of massive stars is crucial to avoid misinterpretations and solution degeneracities. This requires to go beyond photometry-based studies and ideally multi-epoch spectroscopy to quanitatively analyse the components. How much this can change our perception can hardly be underestimated. Table\,\ref{tab:binmis} highlights two recent examples from \citet{Ramachandran+2024} and \citet{Pauli+2022} in the SMC: The inferred values from photometry or spectral-type calibrations change drastically when presumably single objects turn out to be post-interaction binaries, implying that our interpretations of less-studied objects can be significantly blurred. The examples further show that long-predicted binary-evolution products (e.g., intermediate-mass stripped stars, He-burning O stars) actually ``hide'' in what one might consider the ``known'' OB population.

\section{Selected recent discoveries leading us to reconsider our assumptions}

	As illustrated with the discoveries above and further boosted by the new puzzle pieces from Gravitational Wave measurements, reconsidering what we think we know will be an important part towards a more coherent picture of massive stars. While doing injustice to many other important aspects, the proceedings limitations force a limit of the discussion to two examples.

  \subsection{Weak winds: A problem or a matter of diagnostics?}

     While hot, massive stars can show very strong winds in some regimes, their inherent physical dependencies (e.g., on $Z$ and $L/M$, i.e., the proximity to the Eddington limit) can also yield to regimes where the winds are weak \citep[typical definition $\dot{M} < 10^{-8}\,M_\odot\,\text{yr}^{-1}$, see][]{Martins+2005}, even in the Milky Way. As the associated measured mass-loss rates $\dot{M}$ are lower than what is predicted by certain mass-loss predictions, this is also labelled as the \emph{weak-wind problem}. However, many $\dot{M}$-predictions are based on extrapolating a set of explicit model calculations and thus it is by no means clear whether the insufficiency of a prediction also indicates an actual problem in our physical understanding of radiation-driven winds. For example, even within a CAK-type framework, \citet{Kudritzki2002} noted deviations at low metallicity from power-law trends. Moreover, ion decoupling can happen at low wind densities \citep[e.g.,][]{Krticka+2003}. Strong non-linearities in the $\dot{M}(L)$ behaviour are also obtained for the winds of classical WR stars \citep[e.g.,][]{Sander+2020,Moens+2022}, likely leading to extremely weak winds for compact, stripped stars below a certain luminosity thereshold \citep[][though see M{\"u}ller-Horn et al., subm.]{Goetberg+2023}. Hence, there are regimes where we expect $\dot{M}$ to be low and thus have ``genuinely'' weak winds without a ``weak-wind problem''.
					 
Instead, the \emph{weak-wind problem} in its actual meaning deals with stars where $\dot{M}$ derived from standard UV and optical diagnostics is lower than what could theoretically be driven by radiation pressure. This has first been discovered in Galactic late-type O dwarfs \citep{Martins+2005,Marcolino+2009,Lucy2010}, but is suspected to encompass almost the whole range of O dwarfs at SMC metallicity \citep{Ramachandran+2019}. While in a strict sense a detailed, local calculation of the radiative forces and hydrodynamics is necessary to confirm that the stars could drive a stronger wind than observed and thus is subject to the weak-wind problem, an important initial diagnostic is to calculate the \emph{work ratio}
	\begin{equation}
  Q := \frac{\int\left(  a_\text{rad} - \frac{1}{\rho} \frac{\mathrm{d} P}{\mathrm{d} r} \right) \mathrm{d} r}{\int\left( v \frac{\mathrm{d} v}{\mathrm{d} r} + \frac{G M_{\ast}}{r^2} \right) \mathrm{d} r}\text{.}
\end{equation}	 
$Q$ is a global quantity that gives an impression of energy conservation by integrating over the wind-driving forces of radiation and gas pressure in the nominator and the repulling forces of inertia and gravity in the denominator. Assuming all sufficient physics are included, $Q$ should ideally be unity for any stellar atmosphere model resembling an observed spectrum. However, $Q < 1$ is rather the norm. In principle, this could be due to the neglection of elements critical for the radiative driving, but unimportant for the spectral fit. However, while adding further elements in many cases increases the $Q$ value, it often still does not approach unity. Predictions for OB stars from \citet{Bjoerklund+2021} therefore yield lower $\dot{M}$-values, but some of the lowest predictions seem to be in conflict with observations \citep[e.g.,][]{Hawcroft+2021}.

Stars subjected to the weak-wind problem on the other hand have $Q > 1$ with prominent examples like $\beta$\,CMa (B1II-III) or $\mu$\,Col (O9.5V) showing $Q \approx 9$ \citep{Todt+2013}. Initially, the supression of winds by magnetic fields has been discussed as an origin, but this could be ruled out at least for the Galactic targets \citep{Grunhut+2017}. The weak-wind problem in the SMC seems to be less extreme, but persistent. Inspecting the models for SSN\,15 (O5.5V) and SSN\,22 (O6V) from \citet{Rickard+2022} yields $Q \approx 2$. A similar value of $\sim$$2.2$ is found for SSN\,7b (O5.5V) analyzed by \citet{RickardPauli2023}. For later-type O stars, the situation is harder to check as only upper limits were determined in many cases. The \citet{Rickard+2022} models show values close to unity (e.g., 0.93 for SSN\,59 and 0.82 for SSN\,105, both O9.5V) but as the true $\dot{M}$ might be even lower, the weak-wind problem might persist in this regime.
							
For the Galactic case of $\mu$\,Col, \citet{Huenemoerder+2012} presented the intriguing solution that the winds of the stars subjected to weak-wind problem might rather be the problem of a multi-phased wind and missing diagnostics. From X-ray spectroscopy, they derived a six-times higher $\dot{M}$, concluding that a significant fraction of the wind is in a hot phase, invisible to the UV and optical diagnostics. Recently, this idea got new support from JWST mid-IR observations of the nearby O9 dwarf 10\,Lac, where forbidden high-ionization fine-structure emission lines indicate a ten-times higher $\dot{M}$ \citep{Law+2024}. Notably, these lines are not visible in the JWST mid-IR spectrum of $\mu$\,Col, raising the question why two stars with very similar spectral types require so different diagnostics. Nonetheless, it is evident that the treatment of wind structure and mass-loss diagnostics in spectral analysis will have to be revised as there is also independent theoretical demand for this \citep[e.g.,][]{Lagae+2021,Fryer+2025,Moens+2025}. Nonetheless, more observational guidance would be extremely helpful as it is so far unclear how regime-dependent the presence of a hotter wind component is and whether there is a smooth transition from the hot phase actually being a tail of a distribution to presumably becoming a dominant component in stars subjected to the ``weak-wind problem''.

  \subsection{The different roles and faces of Wolf-Rayet stars}

    Wolf-Rayet (WR) stars, at first glance, are the complete opposite of weak-wind stars. Known for their
		strong, prominent emission lines, the winds of these objects can be so dense that the spectrum completely originates in the outer wind. 
		However, this does not apply to all WR stars and the treatment of WR stars
		as one big homogeneous group is another example, where a blurred perception can have striking consequences.
		The spectral classification into WN, WC, and WO with further subtypes defined by line ratios provides a first distinction, but the association with both evolution and feedback is tricky. As the ratio of two strong lines can be the same as the ratio of two weaker lines, objects with very different kinetic feedback as well as different hydrostatic radii can appear under the same label. 
		The association with stellar structure and evolution through the subtypes is therefore non-trivial with many pitfalls.
		
		Strong constraints exist for WC and WO stars, which never show hydrogen (H) in their spectra and thus have to be He-burning (``classical'') WR (cWR) stars. WN stars instead can either be H- or He-burning, unless their outer layers are (almost) depleted in H.  
		Very massive stars (VMS) close to the Eddington limit can appear as H-containing WN (``WNh'') stars already during the main sequence.
		These stars likely lose enough material by themselves to evolve into a cWR, presumbly passing through different WNh subtypes.
		Given sufficient intrinsic mass loss, eventually a hydrogen-free WN stage will be reached. In case of stronger stripping, also former layers
		undergoing He-burning will be revealed and form a C- and O-enriched wind typically reflected in a WC-type spectrum. The transition from a 
		WN to a WC stage is commonly predicted to be extremely fast. However, a small but notable population of WN/WC-transition stars as well as
		WC stars with remaining nitrogen indicate that a considerable transition layer is typical, pointing at a considerable mixing
		above the convective He core \citep{Langer1991}.

    Despite uncertainties such as the amount of He core overshooting the later part of hydrogen-free WR evolution could be considered reasonably understood.
		However, the association with the observed population remains challenging. The bulk of the known WR population actually consists of objects where either the current structure is not in line with evolutionary predictions or the pathway to forming these WR stars is totally unclear. A direct entry into the WR stage has essentialy only been observed between luminous blue variables (LBVs) and late-type WNh stars \citep[e.g.,][]{Maryeva+2019}. The loss of the envelope, in particular below the VMS regime, is a classic explanation in massive single-star evolution \citep{Langer+1994}. However, the association of LBVs and in particular giant eruptions as an outcome of the single-star evolution is debated \citep[e.g.][]{Mehner+2021,Smith2026}. Indeed, binarity is common among WR stars, although this does not imply that binary interaction is always the (main) cause for reaching the observed configurations \citep[e.g.][]{Shenar+2020}. 
		Yet, binary interaction might be an important route to produce certain WR subtypes with strong winds at comparably lower luminosities. For example, the WR population of Wd1 contains mostly WN5-7 and WC8-9 stars with dust excess and X-ray luminosities suggesting all of them to be binaries \citep[][]{Crowther+2006,Anastasopoulou+2024}. Single-star evolution invoking LBV-like stripping at cooler temperatures yield scenarios where these WR subtypes are short-lived as the stars are losing their outer layers and should sit approximately on the He main sequence. Their appearance could be cooler due to strong winds as noted above, but dynamically-consistent atmosphere models only provides reasonable explanations up to WN4 and WC5 \citep[][Lefever et al., subm.]{GraefenerHamann2005} as otherwise the predicted line strengths become too large compared to the observations. A missing link in the formation of mid-type WN binaries could be sgB[e] stars, masking post-common envelope systems. In the LMC, the sgB[e] HD 38489 (aka LHA 120-S134) shows broad He\,II~4686\,\AA-emission within a B[e]-type spectrum \citep[e.g,][]{Massey+2014} and many authors conclude that the B[e] phenomenon does not reflect the underlying object(s), but only shows the circumstellar material \citep[e.g.,][]{Zickgraf+1986,Clark+2013}. Intestingly, the X-ray spectrum of the sgB[e] Wd1-9 also indicates a WR+OB binary \citep{Anastasopoulou+2025}. 
    The later-type WC+O systems can be important cosmic dust sources. Recent JWST observations of more isolated systems revealed long-lasting spiral dust patterns created during the periastron passages \citep[e.g.,][]{Lau+2022,Richardson+2025}.

\begin{figure}[t]
\includegraphics[width=\textwidth]{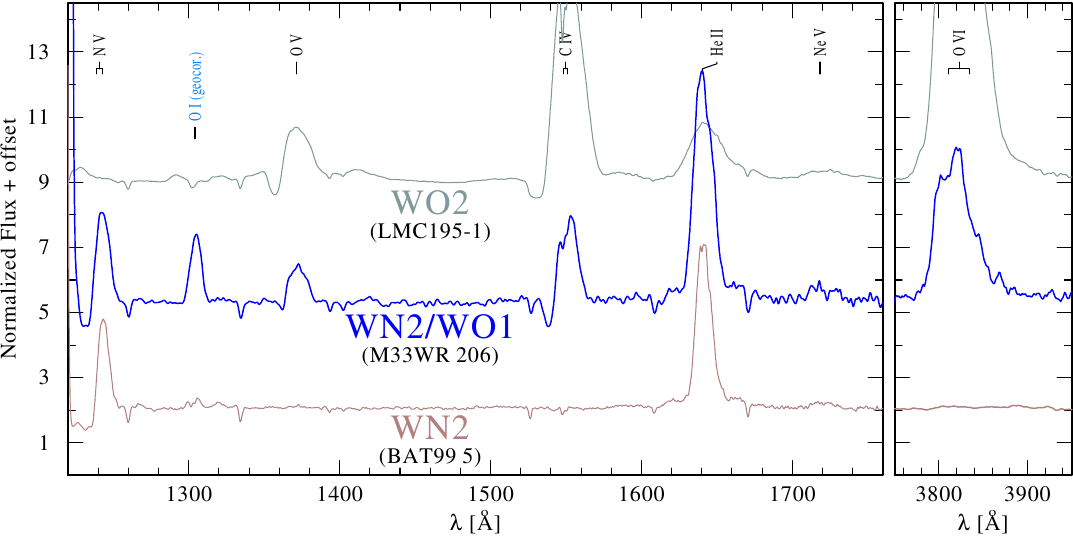}
\caption{Part of the normalized UV and optical spectra from three hot WR stars to demonstrate the direct spectral transition from WN2 to WO. The transition-type object M33WR\,206 \citep[analyzed in][]{Sander+2025} combines the WN2 and WO spectral features while also showing intermediate surface abundances.}
\label{fig:wnwo}
\end{figure}
		
		Contrary, on the very hot end, early-type WN2 and WN3 stars can show comparably weaker winds but then are strong sources of He\,II ionizing flux. 
		In the SMC, half of the WR population is of WN3h-type with the strongest optical emission line (He\,II~4686\,\AA) being just about 
		twice as large as the continuum. This makes these stars extremely easy to hide in integrated light \citep{Gonzalez-Tora+2025}. 
	  While the WN3 stars in the SMC all contain at least some hydrogen, the even hotter WN2 type can be found in the outer regions
		of the Milky Way, M33, and the LMC, likely also being connected to subsolar metallicity. M33 contains a particular interesting WN2 star
		showing also the characteristic strong O\,VI emission typical for WO stars (cf.\ Fig.\,\ref{fig:wnwo}). 
		\citet{Sander+2025} could show that this object is in fact
		in a WN-to-WO transition with another WN/WO star found in M33. This stage
		marks a hotter counterpart of the aforementioned WN/WC stage, happening when the wind mass-loss rates are insufficient to
		considerably cloak the hotter wind-launching layers. Yet, the oxygen abundances of WN/WO stars are considerably lower than their carbon
		abundances, in line with what was found in a recent investigation of LMC WO stars by \citet{Aadland+2022}.
		
	  \begin{table}
			 \centering
			 \caption{\label{tab:ion} Ionizing flux rates of selected hot stellar targets}
		\begin{tabular}{lcccccc}	
			\midrule
				          &  SpT       &  $\log L/L_\odot$  &  $\log Q_\text{H\,I}$  & $\log Q_\text{He\,II}$ & $\log q_\text{He\,II}$  & Source \\\midrule
		 WR\,1        &  WN4       &     $5.6$          &      49.3         &      41.1         &      18.3       &  L26       \\
		 BAT99\,52    &  WC4       &     $5.75$         &      49.5         &      40.4         &      17.4       &  Sa25, t.w.  \\[0.2em]  
		 SMC AB\,10   &  WN3       &     $5.65$         &      49.5         &      48.1         &      24.6       &  \!\!H15, GT25\!\!  \\
		 BAT99\,5     &  WN2       &     $5.5$          &      49.3         &      48.4         &      25.8       &  Sa25        \\
		 M33WR\,206   &  \!\!WN2/WO1\!\!   &     $5.65$         &      49.5         &      48.4         &      25.4       &  Sa25        \\
		 LMC195-1     &  WO2       &     $5.41$         &      49.1         &      48.3         &      25.8       &  A22, t.w.  \\[0.2em]
     CHE 20\,$M_\odot$-pMS &  (WO)      &     $5.67$         &      49.5         &      48.6         &      25.4       &  \!\!K19, Sz25\!\!   \\ 
		 Pop III 100\,$M_\odot$  &   --       &     $6.12$         &      49.9         &      48.7         &      24.6       &  t.w. \\\midrule
		\end{tabular}		
      \tabnote{\textit{Notes}: The two last rows show theoretical models for (a) a $20\,M_\odot$-star undergoing chemically homogeneous evolution at the beginning of core-He burning \citep{Kubatova+2019} and (b) a 100\,$M_\odot$ Pop III model on the zero-age main sequence ($T_\text{eff} \approx 90\,$kK)\\
			\textit{References}:  A22: \citet{Aadland+2022} GT25: \citet{Gonzalez-Tora+2025}, H15: \citet{Hainich+2015}, L26: Lefever et al. (submitted), Sa25: \citet{Sander+2025}, Sz25: \citet{Szecsi+2025} , t.w.: this work} 
		\end{table}

		The WN3h stars and the direct WN/WO transition show that WR stars can be strong sources of He II ionizing flux throughout at least 
		a significant part of their He-burning lifetime. This makes them very interesting sources in the context of observed nebular He\,II
		emission in lower-metallicity galaxies across a wide redshift range. In Table\,\ref{tab:ion}, we list a few examples and compare
		them both early-type WR stars with stronger winds as well as models yielding a similar He\,II ionizing flux from two more exotic
		models commonly invoked to explain observed high ionization: a Pop III star and He-burning object having undergone chemically homogeneous evolution (CHE). 
		The comparison shows that in principle hot WRs with comparably weaker winds could yield the necessary hard ionizing flux. 
		Yet, current evolution and population synthesis models essentially cannot
		reproduce the observed WN2 and WN3h stars, in particular as most of them do not have a companion. For the SMC WN3 stars,
		a detailed multi-epoch monitoring actually also ruled out essentially all plausible dark massive companions \citep{Schootemeijer+2024},
		leaving another open question in massive star research.

\section{Summary: Pathways to answer the many open questions about massive stars}

		Massive stars remain at the forefront of astrophysical research as puzzle pieces on the nature and impact of massive stars emerge at all redshifts. Individual, nearby massive stars are crucial for our detailed physical understanding and still bear many surprising insights. Larger samples and the rigorous studies of young clusters are important for statistical insights and put the life and feedback of massive stars in context. Spectroscopic observations, ideally multi-epoch, are the key to get proper answers, in particular as multiplicity is almost everywhere, but aquiring this data for large samples remains an ongoing challenge. The UV marks a particularly crucial regime here as it bears unique diagnostics for many decisive outcomes of binary interaction.

		Connecting resolved and integrated observations of massive stars is challenging, but essential for a coherent understanding of massive stars and to ensure population synthesis correctly maps the insights from resolved populations. This is particularly crucial as many regimes will remain limited to integrated-light studies. To close knowledge gaps, forthcoming facilities such the ELT or eventually the Habitable World Observatory can become gamechangers, but need to be designed with massive stars in mind to provide the necessary senstivitiy and diagnostics.

%%%%%%%

\begin{acknowledgements}
  AACS acknowledges support by the Deutsche Forschungsgemeinschaft (DFG) in the form of an Emmy Noether Research Group -- Project-ID 445674056 (SA4064/1-1, PI Sander) -- and a Research Grant under Project-ID 496854903 (SA4064/2-1, PI Sander). AACS further acknowledges funding from the Federal Ministry for Economic Affairs and Energy (BMWE) via the Deutsches Zentrum f\"ur Luft- und Raumfahrt (DLR) grants 50 OR 2306 and 50 OR 2503.	This project was co-funded by the European Union (Project 101183150 - OCEANS).
\end{acknowledgements}

\bibliographystyle{iaulike}
\bibliography{sanderbib.bib}

\end{document}